\documentclass[conference]{IEEEtran}
\IEEEoverridecommandlockouts
\usepackage{cite}
\usepackage{amsmath,amssymb,amsfonts}
\usepackage{algorithmic}
 
\usepackage{textcomp}
\usepackage{xcolor}
\usepackage{cite}

\usepackage{booktabs}
\usepackage{hyperref}
\usepackage{caption}
\usepackage[justification=centering]{subcaption}
\usepackage{enumitem}
\usepackage{multirow}
\usepackage{outlines}

\def\BibTeX{{\rm B\kern-.05em{\sc i\kern-.025em b}\kern-.08em
    T\kern-.1667em\lower.7ex\hbox{E}\kern-.125emX}}

\usepackage[linesnumbered]{algorithm2e}
\usepackage{amsthm}
\usepackage{xcolor}
\definecolor{commentgreen}{rgb}{0,0.5,0}

\SetCommentSty{mycommfont}

\usepackage{balance} 

\usepackage{ifsym}

\usepackage{amsmath}
\usepackage{braket}
\usepackage{mathtools}

\usepackage[utf8]{inputenc}
\usepackage[english]{babel}

\theoremstyle{definition}

  \newtheorem{exmp}{Example}[section]

\newcounter{RihanNOC}
\stepcounter{RihanNOC}

\newcommand{\para}[1]{\noindent\textbf{#1.}}
\newcommand{\parab}[1]{\vspace{1mm}\noindent\textbf{#1}} 

 \usepackage{url}

\usepackage{cleveref}
\usepackage{hyperref}

  \usepackage{graphicx}
\begin{document}

\title{Quantum Data Management: From Theory to Opportunities}

\author{\IEEEauthorblockN{Rihan Hai}
\IEEEauthorblockA{\textit{Web Information Systems Group} \\
\textit{ Delft University of Technology}\\
r.hai@tudelft.nl}
\and
\IEEEauthorblockN{Shih-Han Hung}
\IEEEauthorblockA{\textit{Institute of Information Science} \\
\textit{Academia Sinica}\\
shung@cs.utexas.edu }
\and
\IEEEauthorblockN{Sebastian Feld}
\IEEEauthorblockA{\textit{Quantum Machine Learning Group} \\
\textit{ Delft University of Technology}\\
s.feld@tudelft.nl}

}


\maketitle

\begin{abstract}
Quantum computing has emerged as a transformative tool for future data management. Classical problems in database domains, including query optimization, data integration, and transaction management, have recently been addressed using quantum computing techniques. This tutorial aims to establish the theoretical foundation essential for enhancing methodologies and practical implementations in this line of research. 
Moreover, this tutorial takes a forward-looking approach by delving into recent strides in quantum internet technologies and the nonlocality theory.
We aim to shed light on the uncharted territory of future data systems tailored for the quantum internet. 

\end{abstract}


\section{Introduction}

In recent years, the realm of computer science has been abuzz with the potential of quantum mechanics. Quantum computing, which leverages the principles of superposition and entanglement, promises computational capacities far beyond what traditional computers can achieve \cite{nielsen2010quantum}. Moreover, the \emph{quantum internet} holds the potential to offer advantages and capabilities beyond the power of classical internet systems, e.g., secure communication \cite{5438603} or distributed computing \cite{nickerson2014freely}. A quantum internet is a network connecting end nodes that range from simple quantum devices with one qubit to large-scale quantum computers \cite{doi:10.1126/science.aam9288}. Recent studies showed the possibility of real-world quantum internet in the scale of kilometers, e.g., 248 kilometers realized using optical fiber \cite{neumann2022continuous} and 1203 kilometers with satellite \cite{yin2017satellite}.
One vision is that there will be cloud data centers across continents linked by quantum internet, with quantum entanglement promising instantaneous, consistent, and ultimately secure data transmission. 

Since the 1960s, database systems have evolved significantly, transitioning from early hierarchical and network models to the widespread adoption of the relational model \cite{codd1970relational} and relational databases in the 70s-80s. With the birth of the World Wide Web and the surge of web-based applications in the 90s, we witnessed the rise of distributed databases and object-oriented databases \cite{qszu1991principles, atkinson1990object}. In the 2010s, the emergence of big data technologies, NoSQL databases, and cloud computing further reshaped the landscape of data management \cite{abadi2016beckman}. 
As data continues to grow drastically in both volume and variety, the conventional means of data management will eventually hit the ceiling. 
Quantum computing, with its revolutionary potential, positions itself as a pivotal future technology in the ongoing evolution of data management systems.

With the rapid advancement and increasing accessibility of quantum computing technologies \cite{arute2019quantum,zhong2020quantum,wu2021strong,madsen2022quantum}, a pressing need arises to understand the intricacies of data management using quantum computers and the quantum internet. 
Guaranteeing the management and accessibility of data within a quantum computing environment is prominent. 
This requires understanding and utilizing the innovations of quantum technologies. Even further, it calls for a fundamental rethinking of data handling and management paradigms. 
\emph{Entering the era of quantum computing, what is the future of data management?}
\begin{figure}
\centering

 \includegraphics[width=0.97\linewidth]{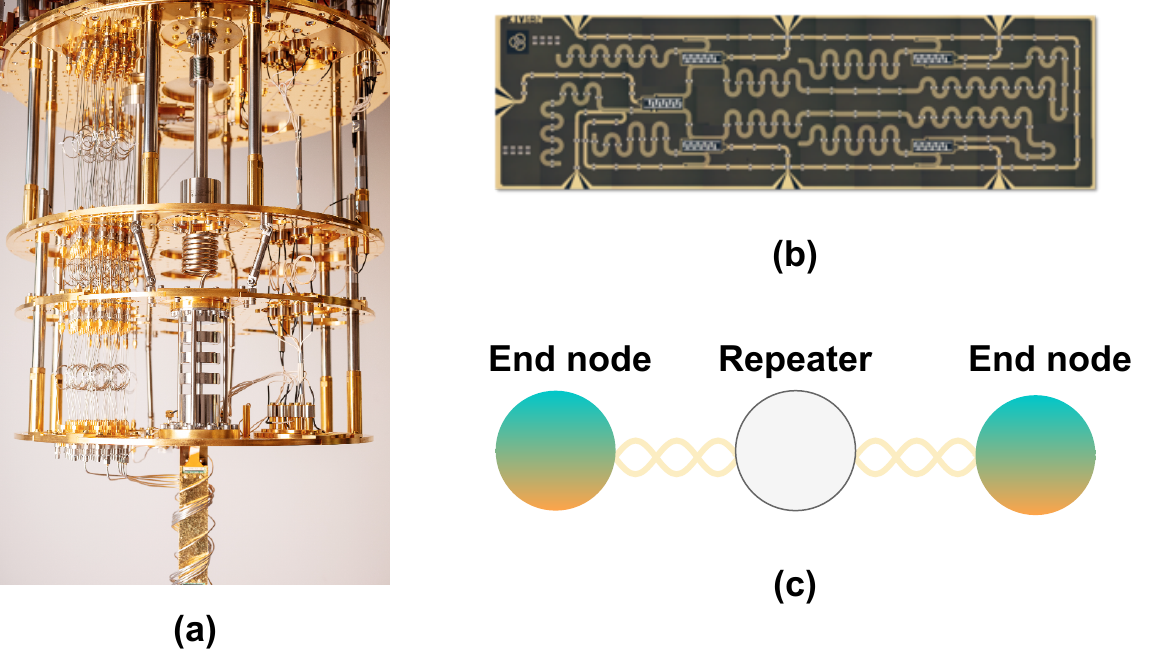} 

\caption{(a) Cloud quantum computer\protect\footnotemark at QuTech, TU Delft.
(b) A five-qubit quantum chip using superconducting technology
\cite{QCimage}. 
(c) The basic unit of a quantum internet: end nodes (quantum devices) and a repeater. The repeater establishes quantum entanglement (yellow lines) with each end node, enabling data transmission through quantum teleportation \cite{5438603}.}
 \vspace{-0.4cm}
\end{figure}
\footnotetext{Quantum Inspire (\url{www.quantum-inspire.com}). Photo taken by Marieke de Lorijn for QuTech.}

\section{Fundamentals of quantum computing}
\label{sec:challenge}
We will start the tutorial by introducing the basic concepts of quantum computing \cite{nielsen2010quantum} related to our discussion in Sec.~\ref{sec:qc}-\ref{sec:qi}. For understandability, we introduce below qubits with their analogous concept in classical computation, i.e., bits. We explain qubits at the abstract mathematical level, such that it is general for diverse software and hardware realization.

\subsection{Basics: qubit, superposition, and entanglement}
\parab{Quantum bit}, or qubit, is the fundamental concept of quantum computing. 
Classical computers use binary digits, bits, as the basic unit of information, which can exist in one of two states: 0 or 1. Similarly, two basic states for a qubit are $\ket{0}$ and $\ket{1}$. To represent the state of a quantum system, the term \emph{ket} is used, denoted with the Dirac notation $\ket{\ }$.

\para{Superposition} The first difference between a classical and quantum bit is: a classical bit can only be 0 or 1, like a coin with a head or tail; but a qubit can exist in a \emph{superposition} of $\ket{0}$ and $\ket{1}$ at the same time.
Mathematically, the superposition of a qubit is represented using the linear combination of basic states $\ket{0}$ and $\ket{1}$: 
\begin{equation*}
\vspace{-0.1cm}
\ket{\psi} = \alpha \ket{0} + \beta \ket{1}  
\end{equation*}

Here, $\ket{\psi}$ is the state of the qubit with $\alpha$ and $\beta$ being complex numbers. 
$\alpha \ket{0} + \beta \ket{1}$ means that a qubit would be measured as 0 with the probability of $\left | \alpha \right |^{2}$ and 1 with the probability of $\left | \beta \right |^{2}$. 
\begin{exmp} Consider the qubit in the following superposition:
	\label{exmp:superposition}
\begin{equation*}
\ket{\psi} =  \frac{1}{\sqrt 2} \ket{0} +  \frac{1}{\sqrt 2} \ket{1}  
\end{equation*}

When measuring this qubit, there is an equal probability of 50\% ($\left | \alpha \right |^{2}=\left | \beta \right |^{2}=\frac{1}{ 2}$) to get a 0 or 1 as a result.
\end{exmp}

\para{Entanglement} Another difference between qubits and bits is \emph{entanglement}~\cite{einstein1935can,schrodinger1935discussion}, referred to as the \emph{spooky action at a distance} by Einstein. 

Imagine two qubits with qubit A being in Amsterdam and qubit B in San Francisco. When qubit A and qubit B are entangled this means their states are correlated. The \emph{spooky} part is that if we measure qubit A in Amsterdam and find it in a state of 0 (or 1), we instantaneously know the state of qubit B is 0 (or 1), even though it is far away in San Francisco. This correlation is maintained independently of the physical distance between them. 


We can consider entangled qubits as a valuable resource in quantum computing \cite{nielsen2010quantum}, which cannot be provided by classical bits. This shared state allows quantum computers to process information in ways that classical computers cannot. 
%
A quantum computer with several entangled qubits in superposition can be processed in a single operation, providing a significant advantage for certain problems, e.g., factoring large numbers \cite{shor1999polynomial} or searching unsorted databases \cite{grover1996fast}.

\begin{table*} [t]
  \centering
  \caption{Recent data management works using quantum computers: an overview}
\label{tbl:DB}
\begin{tabular}{|c|c|c|c|c|c|}
\hline
\textbf{Reference}                                                                              & \textbf{DB problem}                 & \multicolumn{1}{c|}{\textbf{Subproblem}}                                                                           & \textbf{Formulation}  & \multicolumn{1}{c|}{\textbf{\begin{tabular}[c]{@{}c@{}}Intermediate \\ quantum algorithm\end{tabular}}}        & \textbf{Quantum computer}     \\ \hline
\cite{trummer2016multiple}                                                & \multirow{5}{*}{Query optimization} & \multicolumn{1}{c|}{\multirow{2}{*}{\begin{tabular}[c]{@{}c@{}}Multiple query \\ optimization (MQO)\end{tabular}}} & \multirow{4}{*}{QUBO} & \multicolumn{1}{c|}{--}                                 & Annealing-based               \\ \cline{1-1} \cline{5-6} 
\cite{fankhauser2021multiple, fankhauser2023multiple}                     &                                     & \multicolumn{1}{c|}{}                                                                                              &                       & \multicolumn{1}{c|}{QAOA}                                 & Gate-based                    \\ \cline{1-1} \cline{3-3} \cline{5-6} 
\cite{SchonbergerSIGMOD22, schonberger2023SIGMOD, schonberger2023quantum} &                                     & \multicolumn{1}{c|}{\multirow{3}{*}{Join ordering (JO)}}                                                           &                       & \multicolumn{1}{c|}{QAOA}                               & Gate-based \& annealing-based \\ \cline{1-1} \cline{5-6} 
\cite{GroppeBiDEDE23}                                                     &                                     & \multicolumn{1}{c|}{}                                                                                              &                       & \multicolumn{1}{c|}{QAOA, VQE}                               & Gate-based \& annealing-based               \\ \cline{1-1} \cline{4-6} 
\cite{GroppeQMLBiDEDE23}                                                  &                                     & \multicolumn{1}{c|}{}                                                                                              & --                    & \multicolumn{1}{c|}{VQC} & Gate-based                    \\ \hline
\cite{ScherzingerVLDB2023demo}                                            & Data integration                    & \multicolumn{1}{c|}{Schema matching}                                                                               & QUBO                  & \multicolumn{1}{c|}{QAOA}                                 & Gate-based \& annealing-based \\ \hline
\cite{GroppeIDEAS20, OJCC,  groppe2021optimizing}                                                      & Transaction management              & Two-phase locking                                                                                                  & QUBO                  & --                                                      & Annealing-based               \\ \hline
\end{tabular}
 \vspace{-0.3cm}
\end{table*}

\subsection{Quantum computers and libraries} 

\para{Architecture} Quantum computers come in various models, using different approaches to maintain and manipulate the quantum state of qubits, and are suitable for specific tasks. The most important types of quantum computers include quantum annealers and physical systems for implementing gate-based models such as 
trapped-ion quantum computers,
superconducting qubits,
topological quantum computers, 
and photonic quantum computers.  
In the tutorial, we will cover both quantum annealers and gate-based models, which are utilized in existing works (cf. Table~\ref{tbl:DB}).
Quantum annealers are designed to solve optimization problems by finding the lowest energy state of a system, which should correspond to the optimal solution. Quantum annealers can currently be built with more qubits than other existing types of quantum computers, e.g., 5000-qubit D-Wave Advantage quantum annealing processor \cite{king2023quantum}. Gate-based models are universal, i.e., they are in theory capable of performing any computational task that can be executed on a classical computer, but potentially more efficient for certain problems. 

\para{Open-source libraries} 
For developing applications on quantum computers, key resources include open-source toolkits like IBM's Qiskit\footnote{ \url{https://qiskit.org/}}, Microsoft's QDK\footnote{\url{https://learn.microsoft.com/en-us/azure/quantum/overview-what-is-qsharp-and-qdk}}, and Python libraries such as Google's Cirq\footnote{\url{https://github.com/quantumlib/Cirq}}, Amazon's OQpy\footnote{ \url{https://github.com/openqasm/oqpy}}, 
TU Delft's 
OpenQL\footnote{\url{https://openql.readthedocs.io/en/latest/}}, and Qibo\footnote{\url{https://github.com/qiboteam/qibo}}.



%
%


\section{Data management using quantum computers}
\label{sec:qc}

When it comes to quantum computing applied to data management, we are still in the nascent stages. Pioneering efforts have been made in both the theoretical design and early prototyping of database problems for quantum computers, which we briefly explain in Sec.~\ref{ss:early}. 

In Table~\ref{tbl:DB}, more recent works cover essential data management topics such as query optimization \cite{trummer2016multiple, fankhauser2021multiple, fankhauser2023multiple, SchonbergerSIGMOD22, schonberger2023SIGMOD, schonberger2023quantum, GroppeBiDEDE23, GroppeQMLBiDEDE23}, data integration \cite{ScherzingerVLDB2023demo}, and transaction management \cite{GroppeIDEAS20, groppe2021optimizing, OJCC}. 
The general methodology is to map a data management problem to a mathematical formulation solvable by a quantum computer. 
These works, despite solving different problems each, are mostly mapped a so-called \emph{quadratic unconstrained binary optimization} (QUBO) problem.  
For quantum computing, currently, QUBO is one of the most widely applied optimization models \cite{glover2018tutorial}. 
During the tutorial, we will elaborate on the existing attempts summarized in Table~\ref{tbl:DB}. In Sec.~\ref{ssec:queryopt}, we discuss the query optimization problem, which is one of the cornerstones of database research.

\subsection{Quantum database search and query complexity}
\label{ss:early}
A \emph{database} is a collection of data stored electronically in a computer system \cite{abiteboul1995foundations, ramakrishnan2003database}.
In the context of data management, we discuss data complexity, and combined complexity of evaluating database queries \cite{vardi1982complexity, papadimitriou1997complexity}, e.g., conjunctive queries, in terms of the sizes of input database instances and queries.

In the context of quantum computing, a ``quantum database'' is a conceptual framework for processing and searching data using quantum algorithms. These algorithms leverage quantum mechanics to address specific problems more efficiently than classical approaches. 
In this context, a key metric for comparing the computational efficiency of quantum and classical algorithms is \emph{query complexity} \cite{ambainis2018understanding}. 
Consider a database with \(N=2^n\) records, in which each record is identified 
by an \(n\)-bit integer label (e.g., ``010'' for a 3-bit label). The goal is to find a target record \(x_0\) using a binary function \(f: \{0, 1\}^n \rightarrow \{0, 1\}\), where
\vspace{-0.3cm}
		\begin{align*}
		\vspace{-0.6cm}
			\begin{split} 
f(x) =  \begin{cases}1, &\text{if $x=x_0$ } \\0, & \text{otherwise}\end{cases}
			\end{split}
 			\vspace{-0.2cm}
		\end{align*}
A quantum algorithm's efficiency can be measured by its query complexity, i.e., how many times it queries this function \(f(x)\). 
One of the most well-known algorithms demonstrating quantum speedup is \emph{Grover's algorithm} \cite{grover1996fast}.
To search a specific record in an unsorted database of \(N\) records, classical algorithms require \(O(N)\) operations, while Grover's algorithm achieves this in \(O(\sqrt{N})\) operations. 
Grover's algorithm has inspired early research into quantum algorithms for searching databases \cite{terhal1998single, boyer1998tight, patel2001quantum, tsai2002quantum, imre2004generalized, ju2007quantum}. 
Moreover, efforts have been made to  develop quantum query languages akin to SQL for classical databases,
covering basic operations:  
join and set operations 
(intersection, union, difference) \cite{cockshott1997quantum, gueddana2010optimized, salman2012quantum, pang2013quantum, gueddana2014cnot, joczik2020quantum}, and data manipulation operations (insert, update, delete)  \cite{gueddana2010optimized, younes2013database, gueddana2014cnot}.

\subsection{Query optimization}
\label{ssec:queryopt}
Recent quantum-related works mainly tackled two subproblems of query optimization: multiple query optimization (MQO) and join ordering (JO). 

\textbf{Multiple query optimization} studies how to choose query plans given a set of queries \cite{sellis1988multiple}, which is NP-hard. As one of the earliest attempts to study the MQO problem with modern quantum computers, 
Trummer et al. proposed in 2016 a method using the D-Wave 2X annealing-based quantum computer \cite{trummer2016multiple}. The proposed method includes logical and physical mapping. At the logical level, it transforms the MQO problem to a QUBO formulation, guaranteeing each query has one query plan selected while minimizing the time-wise efficiency. At the physical level, the logical formula of QUBO is further transformed to the energy formula coherent with the physical design of the quantum computer \cite{mcgeoch2013experimental}. 
The experiments demonstrated 1000x speedup on quantum annealer compared to state-of-the-art MQO solutions at that time, although only for a limited subset of MQO problems. 
Annealing-based quantum computers can reach a larger number of qubits. However, the gate-based model is a more general architecture. Subsequent works \cite{fankhauser2021multiple, fankhauser2023multiple} tackled the MQO problem on gate-based models using the Quantum Approximate Optimization Algorithm (QAOA) \cite{farhi2014quantum}, which is a hybrid quantum-classical algorithm.

\textbf{Join ordering} studies how to identify the optimal ordering of join operations between relations for an efficient query plan  \cite{steinbrunn1997heuristic, DBLP:journals/pvldb/LeisGMBK015, trummer2017solving}. To find optimal left-deep join trees, Sch{\"o}nberger et al. \cite{SchonbergerSIGMOD22, schonberger2023SIGMOD} defined the intermediate formulation based on mixed integer linear programming (MILP), followed by binary integer linear programming (BILP) models, which are eventually transformed to QUBO. 
Recent works also expanded the problem to bushy join trees \cite{schonberger2023quantum, GroppeBiDEDE23}. In \cite{GroppeQMLBiDEDE23}, Winker et al. provided a different angle: treating join ordering as a \emph{learning} problem, then apply quantum machine learning techniques based on a variational quantum circuit (VQC) algorithm \cite{chen2020variational}.


\begin{figure}[t]
    \centering
    \includegraphics[width=0.8\linewidth]{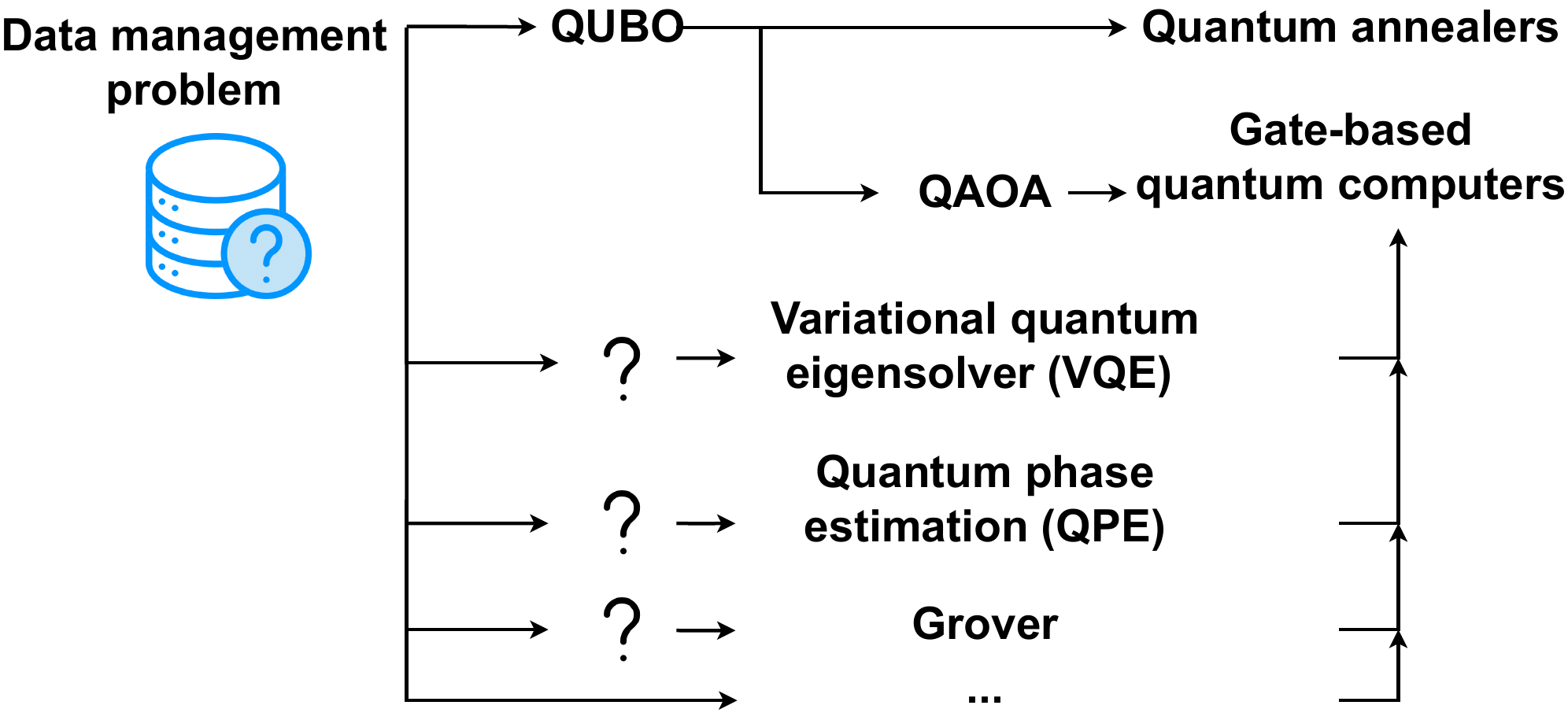}
    \caption{Roadmap for solving data management problems on quantum computers.}
    \label{fig:roadmap}
\vspace{-0.4cm}
\end{figure}

\subsection{Opportunities for data management using quantum computers} 
We outline a few near-term research directions in the following. 
\\ \underline{1) \emph{Database problem reformulation.}} Illustrated in Fig.~\ref{fig:roadmap}, the general methodology to solve data management problems on quantum computers includes two steps. First, we need to reformulate the data management problem (e.g., MQO or schema matching) to a problem which can be solved on quantum computers (e.g., QUBO). 
Besides QAOA, there are many more algorithms for quantum computers, e.g., variational quantum eigensolver (VQE), quantum phase estimation (QPE), and Grover's algorithm. Therefore, the main research opportunities are to explore \emph{which} data management problems can be formalized into problems solvable by quantum computers and provide better or faster solutions compared to classical solutions.
\\\underline{2) \emph{Hybrid approach using classical and quantum computers.}} Most works in Table~\ref{tbl:DB}
require additional steps on classical computers. For instance, in \cite{trummer2016multiple} a prior step is to cluster the queries sharing intermediate results, which significantly reduces the required number of qubits. Many research challenges emerge regarding how to design a hybrid approach using classical and quantum computers. 
\\ \underline{3) \emph{Optimization given quantum computer constraints.}} Despite the compelling computation power of quantum computers, to design data management systems that make use of quantum computers, we still face many practical constraints such as the restricted number of qubits as well as noisy operations \cite{clerk2010introduction}. For a foreseeable period of time, such limitations will coexist while developing data management systems that utilize quantum computers. 
It poses technical challenges and research opportunities to achieve optimal performance (e.g., time-wise efficiency, accuracy, or completeness of the results), given the physical properties and constraints of quantum computers. 
\section{Opportunities: Data Management via Quantum Internet}
\label{sec:qi}
Quantum internet promises unprecedented new capabilities beyond our internet today \cite{kozlowski2019towards}. The applications include distributed systems \cite{doi:10.1126/science.aam9288}, quantum cloud computing \cite{castelvecchi2017ibm}, quantum key distribution \cite{bennett2014quantum}. One of the most fundamental theories enabling these new capabilities is \emph{quantum nonlocality} \cite{RevModPhys.82.665}. 
Intuitively, we can understand quantum nonlocality as: in a distributed system, we have two computers in different locations; when we change a file on one computer, the corresponding file on the other computer would change immediately. It bypasses our usual assumptions about network distance and speed limitations to transfer data. 

The study of quantum nonlocality is at subatomic level, i.e., entangled particles. During the tutorial, we will explain representative nonlocality problems (e.g., CHSH \cite{clauser1969proposed} and GHZ \cite{greenberger1989going} games) using logical operators, e.g., AND ($\wedge $), OR ($\vee$). This shares a unified mathematical foundation with data management problems. For example, for query optimization, a query plan can be expressed as relation predicates connected by logical operators \cite{kastrati2018generating}.


\subsection{Quantum nonlocality} 
In Sec.~\ref{sec:challenge}, we mentioned superposition and entanglement. 
To explain the phenomenon of nonlocality, 
we introduce one of the most well-known states for entanglement, \emph{Bell's state}.

\begin{exmp} (\textbf{Bell's state}) We extend the one-qubit state in Example~\ref{exmp:superposition} to a two-qubit state as follows:  
	\label{exmp:EPR pair}
\begin{equation*}
\ket{\Psi}=\frac{1}{\sqrt 2}(\ket{00}+\ket{11}) 
\end{equation*}

\end{exmp}

\para{Nonlocal games} To make it simple, we consider only two players, Alice and Bob. 
A two-player nonlocal game is a game that consists of two players and a referee. \
During the game, each player receives an input (like a question or a number) from the referee. Two players, Alice and Bob, however, cannot communicate once the game starts. 
The intriguing aspect of nonlocal games is that if Alice and Bob share an entangled quantum state, e.g., Bell's state from Example~\ref{exmp:EPR pair}, they sometimes can win the game with a higher probability than classical strategies. We give an example:

\begin{exmp} (\textbf{Clauser-Horne-Shimony-Holt (CHSH) game} \cite{clauser1969proposed}) 
In this game, Alice receives input $x$ and outputs one-bit answers $a$, while Bob receives $y$ and outputs $b$. 
The players win the game if the questions and the answers satisfy $x\wedge y=a\oplus b$, where $\wedge$ is the logical AND operator and $\oplus$ is XOR. 
We omit the calculations here, but show results. 
The two players win optimally with score $\approx 0.85$ using an entangled Bell's state, and every pair of players who do not share entangled states can succeed with probability of at most $0.75$. \ 
\end{exmp}

The CHSH game demonstrates an example of entanglement winning over classical strategies by a higher success probability (0.85 vs. 0.75). During the tutorial, we will give another example, the three-player Greenberger-Horne-Zeilinger (GHZ) game \cite{greenberger1989going}. In the GHZ game, the entangled state achieves a probability of 1, while classical resources can only achieve a probability of 0.75. That is, with entanglement, we can achieve a task that is \emph{not} possible with classical resources.

\subsection{Future research directions of data management via quantum internet}
Quantum nonlocality serves as the theoretical foundation of protocols for secure communication and key distribution \cite{bennett2014quantum}. Naturally, it is an interesting direction to explore secure data management via quantum internet.
Next, we discuss new design challenges and research opportunities, which are not discussed in the existing literature.
\\ \underline{1) \emph{New data structures.}} We face the fundamental challenge of data representation, since communicating quantum information is fundamentally different from classical information. Take the famous \emph{no-cloning theorem} \cite{nielsen2010quantum} for example. Existing database theory and systems are built upon the fact that we can freely read and copy data, e.g., duplicate a data tuple, or replicate a dataset (e.g., a CSV file) on another computer in the network. However, the no-cloning theorem states that we cannot make an exact copy of an arbitrary quantum state. 
Thus: How to design data models, when quantum data cannot be copied without destroying the original version?
\\ \underline{2) \emph{New architectures of distributed data management systems.}} Distributed database systems traditionally manage distributed, fragmented, and replicated data over classical computer networks. Facing quantum properties such as entanglement, nonlocality, and no-cloning theorem, we need new system architectures to ensure reliability, availability, and scalability. The question arises, for instance, how to design distributed data management systems based on protocols used in quantum internet \cite{doi:10.1126/science.aam9288}? 
 Given the quantum hardware or network failures \cite{dulek2020secure}, how to design fault tolerance and recovery strategies? 


\section{Presenters}
\textbf{Rihan Hai}
is an assistant professor at TU Delft, The Netherlands. Her research focuses on data lakes, data management for machine learning, and quantum data management. 

\textbf{Shih-Han Hung} is a postdoc at Academia Sinica. His research aims to better understand the power and the limit of quantum computers. 

\textbf{Sebastian Feld} is an assistant professor at 
TU Delft, The Netherlands. He and
his group are working on Quantum Machine Learning.


\balance
\newpage
\bibliographystyle{IEEEtran}
\bibliography{mybib}

\end{document}